\input{aipcheck}

\documentclass[
    ,final            % use final for the camera ready runs
  ]
  {aipproc}

\layoutstyle{8x11double}

\newcommand{\be}{\begin{equation}}
\newcommand{\ee}{\end{equation}}
\newcommand{\bea}{\begin{eqnarray}}
\newcommand{\eea}{\end{eqnarray}}
\newcommand{\vep}{\varepsilon}

\begin{document}

\title{$U(1)_R$ mediation from the flux compactification \\ in six dimensions}


\author{Hyun Min Lee}{
  address={
Department of Physics,
Carnegie Mellon University, Pittsburgh, PA 15213, USA}
}

\begin{abstract}
 We consider a supersymmetric completion of codimension-two branes with nonzero tension
 in a 6D gauged supergravity. 
 As a consequence, we obtain the football solution with 4D Minkowski space as a new supersymmetric background that preserves 4D ${\cal N}=1$ SUSY.
 In the presence of brane multiplets, we derive the 4D effective supergravity action
 for the football background and show that the remaining modulus can be stabilized by a bulk non-perturbative
 correction with brane uplifting potentials at a zero vacuum energy. We find that the $U(1)_R$ mediation can be a dominant source of SUSY breaking for a brane scalar with nonzero $R$ charge.
\end{abstract}

\maketitle

%%%%%%%%%%%%%%%%%%%%%%%%%%%%%%%%%%%%%%%%%%%%
%% MAINMATTER
%%%%%%%%%%%%%%%%%%%%%%%%%%%%%%%%%%%%%%%%%%%%

Weak-scale supersymmetry(SUSY)\cite{susy}, as a solution to the hierarchy problem, has been considered as one of the leading candidates beyond the Standard Model.
When SUSY is broken at the weak scale, however, generic soft mass parameters would lead to unacceptably large flavor and/or CP violations.
In 4D gravity mediation scenario, there are contact terms between
the visible and hidden sectors,
$$
\Big[\frac{c_{ij}}{M^2_P}Q^\dagger_i Q_j Q^{'\dagger}Q'\Big]_D,
$$
where $Q_i(i=1,2,3),Q'$ are visible and hidden brane fields.
Then, after hidden sector SUSY breaking($F_{Q'}\neq 0$), flavor violating soft masses would be generated as $m^2_{ij}=c_{ij}m^2_0$ for $c_{ij}\neq \delta_{ij}$.

In higher dimensions, on the other hand, the hidden and visible sectors can be localized on different branes so that there would not be direct contact terms, which is the sequestering mechanism\cite{sequester}. 
The idea has been first realized in 5D supergravity on $S^1/Z_2$ for which the K\"ahler potential is
derived\cite{sequester} as
$K=-3\ln (T+T^\dagger-2Q^\dagger Q-2Q^{'\dagger}Q')$
where $T$ is the radion chiral multiplet. 
In view of the superconformal factor $\Omega=-3 e^{-K/3}$, 
the visible sector does not have a direct contact term to either the hidden sector or the radion multiplet. 
It has been shown\cite{luty} that even after the radion is stabilized by brane and bulk non-perturbative corrections, the soft mass of the visible scalar vanishes at tree level.
Thus, anomaly mediation\cite{sequester,anomalymed} could be a dominant source for nonzero soft masses in the visible sector.
However, due to the negative slepton mass problem, one needs to rely on the other mechanism of SUSY mediation. 

In $D>5$ dimensions, the generic K\"ahler potential is not a sequestered form\cite{highseq} as
$K=-p\ln (T+T^\dagger-2Q^\dagger Q-2Q^{'\dagger}Q')$ with $p\neq 3$.
So, there exist direct couplings of the visible sector to
both the hidden sector and the modulus multiplet.
For $p=1$, a 6D ungauged supergravity compactified 
on an orbifold\footnote{The dilaton $S$ and the shape modulus $\tau$ can be stabilized by a modular invariant superpotential given for bulk gaugino condensates.} $T^2/Z_2$ has been considered\cite{fll}. 
In this case, when $\partial_T W= 0$ for which the $T$ modulus is unfixed, a brane scalar soft mass vanishes at tree level\cite{fll}. But, once the $T$ modulus is stabilized by a $T$-dependent superpotential, the brane scalar soft mass does not vanish due to modulus mediation\cite{fll}.
 
In this talk, we consider a flux compactification in a 6D gauged supergravity with codimension-two branes. 
The $U(1)_R$ gauge flux in the internal dimensions is turned on 
to stabilize the $T$ modulus as well as the shape modulus.
Salam and Sezgin\cite{SS} obtained a unwarped solution $M_4\times S^2$ for which 4D ${\cal N}=1$ SUSY remains and only the dilaton $S$ is left unfixed.
We show that 4D ${\cal N}=1$ SUSY is preserved under the deformation of $S^2$ to a football geometry 
with nonzero brane tensions\cite{leepa}. 
In the presence of brane multiplets on the football\cite{lee}, we derive the 4D effective
supergravity with bulk and brane light modes, and find that after the $S$ modulus is also stabilized by bulk gaugino condensates below the compactification scale, the $U(1)_R$ mediation\footnote{See Ref.~\cite{4Dcase} for the case in 4D supergravity.} is the dominant source of the
SUSY breaking for the visible brane scalar with nonzero $R$ charge.

The 6D chiral gauged supergravity\cite{NS} is composed of a gravity multiplet($e^A_M,\psi_M,B^+_{MN}$), and a tensor multiplet($\phi,\chi,B^-_{MN}$) as well as a vector multiplet($A_M,\lambda$), which gauges the $U(1)_R$ symmetry.
Then, 6D anomalies can be cancelled for $244+n_V-n_H=0$ where $n_V,n_H$ are the numbers of vector and hyper multiplets, respectively. The simplest possibility of the anomaly cancellation with $U(1)_R$ is to introduce only $n_H=245$ hyper multiplets containing neutral hyperinos. 

The Salam-Sezgin solution has been generalized to the unwarped or warped 4D Minkowski solutions with nonzero
brane tensions\cite{branesol}. The Lagrangian for a brane tension, ${\cal L}_{\rm brane}=-e_4 T\delta^2(y)$, however, breaks the bulk SUSY explicitly as $\delta {\cal L}_{\rm brane}=-e_4 \frac{1}{4}T({\bar\psi}_\mu \Gamma^\mu\vep+{\rm h.c.})\delta^2(y)$. In the gauged supergravity, as the gravitino is charged under $U(1)_R$, varying
the gravitino kinetic term gives rise to a piece of the gauge field strength as
$$
\delta{\cal L}_{\rm gravitino}=-\frac{i}{2}e_6\, g{\bar\psi}_M\Gamma^{MNP}\vep F_{NP}+\cdots.
$$  
In the above, we rewrite the gauge field strength in terms of the hatted one 
and a localized Fayet-Ilioupolos(FI) term parametrized by $\xi=\frac{T}{4g}$ as\cite{leepa}
\be
{\hat F}_{mn}=F_{mn}-\xi\epsilon_{mn}\frac{\delta^2(y)}{e_2}
\ee
where $\epsilon_{mn}$ is the 2D volume form.
Then, after a $Z_2$ orbifold projection of half the bulk SUSY on the brane
with $\vep_R(y=0)=0$, we can cancel the brane tension term by the variation of the gravitino kinetic term. 
The strength tensors for the gauge field and the KR field appearing in the bulk action and the fermionic SUSY transformations are replaced with the modified ones\cite{leepa}.

Even in the presence of the FI term, we maintain the general warped solutions\cite{leepa}. 
In particular, for a constant dilaton and ${\hat F}_{\rho\theta}=4g\epsilon_{\rho\theta}$, the football solution is obtained as 
\be
ds^2=\eta_{\mu\nu}dx^\mu dx^\nu+\frac{r^2_0}{4}(d\rho^2+\lambda^2\sin^2\rho d\theta^2) 
\ee
with $r^2_0=\frac{1}{2g^2}$. Two codimension-two branes with equal tensions 
$T_1=T_2=4\pi (1-\lambda)$ are located at the poles of the football. 
The gauge field strength is modified due to the localized FI terms as $A_\theta=-\frac{\lambda}{2g}(\cos\rho\mp 1)\pm \frac{\xi_1}{2\pi}$ with $\xi_1=\frac{T_1}{4g}$.
Then, the flux quantization condition imposes
the monopole number $n=1$ while $\lambda$ can be arbitrary. 
It has been shown that
the football solution preserves 4D ${\cal N}=1$ SUSY due
to the localized FI terms\cite{leepa}. 

Brane multiplets can be also accommodated by modifying further the field strength tensors and the SUSY transformations\cite{lee}.
For a chiral multiplet, a brane scalar with $R$ charge $r_Q$ has a mass term, ${\cal L}_{\rm mass}=-4r_Qg^2|Q|^2e_4$, while
the fermionic partner of the brane scalar having an $R$ charge $r_Q-1$ is massless.
The gauge field strength picks up an additional correction,
\be
{\hat F}_{mn}=F_{mn}-(\xi+r_Q g|Q|^2)\epsilon_{mn}\frac{\delta^2(y)}{e_2}.
\ee
The kinetic term for the brane chiral multiplet has a dilaton coupling as ${\cal L}_{\rm kin}=-e_4 e^{\frac{1}{2}\phi}(D^\mu Q)^\dagger D_\mu Q+\cdots$
while the kinetic term for a brane vector multiplet does not depend on the moduli.
Moreover, the brane $F$ and $D$ terms are ${\cal L}_F=-e_4 e^{\psi-\frac{1}{2}\phi}|F_Q|^2$ (with $F_Q=\frac{\partial W}{\partial Q}$ for a moduli-independent brane superpotential $W$, and $e^\psi$ the volume modulus) and ${\cal L}_D=-e_4\frac{1}{2}e^\phi D^2$, respectively. 

Now we turn to the low energy supergravity with brane multiplets for the football geometry.
To that purpose, we take the ansatz for the 6D solution as
\bea
ds^2&=& e^{-\psi(x)}g_{\mu\nu}(x)dx^\mu dx^\nu + e^{\psi(x)} ds^2_2, \nonumber \\
\phi&=&f(x), \ \ \ \
{\hat F}_{MN}=\langle {\hat F}_{MN}\rangle+{\cal F}_{MN},    
\eea
where $\langle {\hat F}_{MN}\rangle$, ${\cal F}_{MN}$ are the VEV and fluctuation of the gauge field strength,
respectively, and $ds^2_2$ is the 2D metric of the football solution.
Then, by solving the 6D equations and the Bianchi identities for the modified field strengths\cite{lee}, 
we obtain 
\bea
{\hat G}_{\mu mn}&=&\Big(-b+4gA_\mu+\frac{J_\mu}{V}\Big)\epsilon_{mn}, \\
{\hat F}_{mn}&=&\Big(4g-\frac{r_Qg|Q|^2}{V}\Big)\epsilon_{mn},
\eea
where $b=-\frac{1}{2}{\cal B}_{mn}\epsilon^{mn}$ for the globally well-defined 
${\cal B}=B-\frac{1}{2}\langle A\rangle\wedge {\cal A}$ that satisfies 
$\delta_{\Lambda_0} (d{\cal B})$=0 for the background gauge transform $\Lambda_0$, 
$J_\mu$ corresponds to the Noether current for brane multiplets,
and $V$ is the volume of extra dimensions for the football solution.
After plugging the above solutions into the 6D action together with $e^f G_{\mu\nu\rho}=\epsilon_{\mu\nu\rho\tau}\partial^\tau\sigma$ 
and integrating over the extra dimensions,
we identify the K\"ahler potential as\cite{lee}
\bea
K&=&-\ln\Big(\frac{1}{2}(S+S^\dagger)\Big)-\frac{2\xi_R}{M^2_P} V_R \\
&&-\ln\Big(\frac{1}{2}(T+T^\dagger-\delta_{GS} V_R)-\frac{1}{M^2_P}Q^\dagger e^{-2r_Qg_R V_R} Q\Big)
\nonumber
\eea
where $\delta_{GS}=8g_R$ and $\xi_R=2g_RM^2_P$  with $g_R=g/\sqrt{V}$ and 
the scalar components of the moduli superfields $S,T$ are given by
$$
S=e^{\psi+\frac{1}{2}f}+i\sigma, \quad T=e^{\psi-\frac{1}{2}f}+\frac{1}{M^2_P}|Q|^2+ib
$$
Here $V_R$ is the $U(1)_R$ vector superfield and $Q$ is the brane chiral superfield.
On the other hand, the gauge kinetic functions for the bulk and brane vector multiplets are
$f_R= S$ and $f_W = 1$, respectively. 

The 4D effective scalar potential is given by the $U(1)_R$ D-term as
\be
V_0=\frac{2g^2_RM^4_P}{{\rm Re}(S)}\bigg[1-\frac{1-\frac{r_Q}{2M^2_P}|Q|^2}{{\rm Re}(T)-|Q|^2/M^2_P}\bigg]^2.
\ee
So, ${\rm Re}(T)=1$ and $|Q|=0$ at the SUSY minimum with a zero vacuum energy while ${\rm Re}(S)$ is undetermined.
Then, the effective brane scalar mass vanishes due to the cancellation between the brane mass term
and the flux-induced mass term. 

In order to stabilize the $S$ modulus, we assume that the bulk gaugino condensates generate an $S$-dependent superpotential $W(S)$. For instance, the double gaugino condensates would lead to a racetrack form,
$W(S)= \Lambda_1 e^{-\beta_1 S}+\Lambda_2 e^{-\beta_2 S}$. 
We denote the resulting additional potential by $V_1=e^K(|D_SW|^2 K^{-1}_{SS^\dagger}-2|W|^2)/M^2_P$. 
For $|\beta_1-\beta_2|\ll \beta_1$, the potential is minimized at a large ${\rm Re} (S)$.
After the $S$ modulus is stabilized, there appears a negative vacuum energy, which needs to be lifted up to a small positive vacuum energy by means of the $F$ and/or $D$ terms on the hidden brane.
Including the non-perturbative correction and the uplifting potentials, 
the 4D scalar potential becomes
\be
V_{\rm tot}=V_0+V_1+V_2+V_3
\ee
with $V_2=\frac{1}{{\rm Re}(S)}|F_{Q'}|^2$ and $V_3=\frac{D^2}{2({\rm Re}(T)-|Q|/M^2_P)^2}$.
Then, $|Q|=0$ is still the minimum for $r_Q({\rm Re}(T)-1)>0$, while the minimum of ${\rm Re}(T)$ is shifted to
\be
{\rm Re}(T)=\frac{1+\frac{1}{2}\alpha D^2}{1-\frac{1}{2}\alpha {\rm Re} (T) V_1}; \quad \alpha\equiv\frac{{\rm Re}(S)}{2g^2_R M^4_P}.
\ee
The $S$ modulus is also determined approximately by $F_S=0$ although it is shifted a bit 
by the brane F-term.

After fixing all the moduli at the zero vacuum energy, we find that the $U(1)_R$ D-term is the only source
of SUSY breaking at tree level for a brane scalar. The soft brane scalar mass is given by
\bea
m^2_Q&=&\frac{V_{Q^\dagger Q}}{K_{Q^\dagger Q}}\bigg|_{Q=0}=r_Q\,g_R D_R|_{Q=0} \nonumber \\
&=& \frac{D^2+{\rm Re}(T)V_1}{1-\frac{1}{2}\alpha {\rm Re}(T)V_1}\frac{\frac{1}{2}r_Q}{{\rm Re}(T)M^2_P}.\label{gscalarmass}
\eea
A brane scalar with $r_Q=0$ has a vanishing mass at tree level.
Since the $T$ modulus is stabilized by the $U(1)_R$ D-term, 
the functional form of the scalar potential is $V=V({\rm Re}(T)-|Q|^2/M^2_P)$ for $r_Q=0$ so it is understood that $\partial_T V=0$ gives rise to $V_{Q^\dagger Q}=0$.
When the gravitino mass is much smaller than the mass of an anomalous $U(1)$ gauge boson relevant for the brane D-term, the F-term uplifting is dominant\cite{foverd}.
For example, we assume that the uplifting potential is given solely by the F-term with 
$W=W(S)+\mu^2 Q'+W_v(Q)$, i.e. $F_{Q'}=\mu^2$. 
Then, for the vanishing vacuum energy condition $|F_{Q'}|^2\simeq -{\rm Re}(S){\rm Re}(T)V_1={\rm Re}(S){\rm Re}(T)(2m^2_{3/2}M^2_P-|F_S|^2)$ with $\alpha {\rm Re}(T) V_1\ll 1$, 
the brane scalar mass becomes 
$m^2_Q\simeq -r_Q\Big(m^2_{3/2}-\frac{1}{2}\frac{|F_S|^2}{M^2_P}\Big)$.
So, for a negative $R$ charge of the brane scalar, the $U(1)_R$ mediation may be rendered dominant over
anomaly mediation for solving the slepton mass problem.
On the other hand, when the SM gauge fields are localized on the brane, 
there is no tree-level gaugino mass due to their trival gauge kinetic functions.
Then, the gaugino masses are given by anomaly mediation so they are loop suppressed compared to
the $U(1)_R$ mediated soft scalar masses.

\begin{theacknowledgments}
  The work is based on the talk delivered in PASCOS-2008 at Perimeter Institute, Canada, and SUSY-2008 at Seoul, Korea, and is to appear in the proceedings of SUSY-2008. 
  The author is supported by the DOE Contracts DOE-ER-40682-143 and
DEAC02-6CH03000.
\end{theacknowledgments}

\end{document}